\documentclass[iop,apj]{emulateapj}
\usepackage{amsmath}
\usepackage{amssymb}
\usepackage{apjfonts}
\usepackage{natbib}
\usepackage{graphicx}
\usepackage{ifpdf}
\usepackage{natbib}
\usepackage{float}
\usepackage{subdepth}

\newcommand\lambdap{\lambda_{\rm p}}
\newcommand\omegap{\omega_{\rm p}}

\begin{document}
\title{The effect of cooling on particle trajectories and acceleration\\ in relativistic magnetic reconnection}

\author{Daniel Kagan\altaffilmark{1,2},  Ehud Nakar\altaffilmark{1},Tsvi Piran \altaffilmark{2}}
\email{daniel.kagan@mail.huji.ac.il}
\altaffiltext{1}{Raymond and Beverly Sackler School of Physics \& Astronomy,  Tel Aviv University, Tel Aviv 69978, Israel}
\altaffiltext{2}{Racah Institute of Physics, The Hebrew University, Jerusalem 91904, Israel}

\begin{abstract}The maximum synchrotron burnoff limit of 160 MeV represents a fundamental limit to radiation resulting from electromagnetic particle acceleration in one-zone ideal plasmas.  In magnetic reconnection, however, particle acceleration and radiation are decoupled because the electric field is larger than the magnetic field in the diffusion region. We carry out two-dimensional particle-in-cell simulations to determine the extent to which magnetic reconnection can produce synchrotron radiation above the burnoff limit. We use the test particle comparison (TPC) method to isolate the effects of cooling by comparing the trajectories and acceleration efficiencies of test particles incident on such a reconnection region with and without cooling them. We find that the cooled and uncooled particle trajectories are typically similar during acceleration in the reconnection region, and derive an effective limit on particle acceleration that is inversely proportional to the average magnetic field experienced by the particle during acceleration. Using the calculated distribution of this average magnetic field as a function of uncooled final particle energy, we find analytically that cooling does not affect power-law particle energy spectra except at energies far above the synchrotron burnoff limit. Finally, we compare fully cooled and uncooled simulations of reconnection, confirming that the synchrotron burnoff limit does not produce a cutoff in the particle energy spectrum.  Our results indicate that the TPC method accurately predicts the effects of cooling on particle acceleration in relativistic reconnection, and that even far above the burnoff limit, the synchrotron energy of radiation produced in reconnection is not limited by cooling.
\end{abstract}
\keywords{magnetic reconnection -- acceleration of particles -- relativity -- radiation mechanisms: non-thermal}

\section{Introduction}

Magnetic reconnection is a process in which magnetic energy is converted into kinetic energy as the topology of the magnetic field changes. This process can produce fast energy transfer and accelerate particles to high energies. In the relativistic reconnection regime (see \citet{kagan_15} for a review) where the ratio of the magnetic energy to the total enthalpy of the particles (the magnetization $\sigma$) is much larger than 1, emission from these high-energy particles has been hypothesized to produce observed high-energy emission in various astrophysical contexts. Systems whose emission may be explained by relativistic reconnection include the prompt phase of gamma-ray bursts (GRBs) \citep{thompson_94, lyutikov_03,giannios_05,lyutikov_06,icmart_11, mckinney_12,zhang_14,beniamini_13, beniamini_16} pulsar wind nebulae \citep{kirk_03,  sironi_11,petri_12}, and active galactic nucleus (AGN) jets \citep{2009MNRAS.395L..29G,giannios_13,nalewajko_11,narayan_12}.

One of the features of magnetic reconnection is that it may be able to evade a constraint on particle acceleration called the synchrotron burnoff limit \citep{guilbert_83, dejager_96}. This constraint can be derived by equating the synchrotron energy loss from a particle (which increases with the Lorentz factor of a particle as $\gamma^2$), with the energy gain from electrical acceleration (which is independent of $\gamma$). This produces a maximum energy reachable by synchrotron radiation from electromagnetically accelerated particles of \begin{equation}\epsilon_{\rm bo, 0}\equiv\frac{9m_e c^2}{4\alpha_{\rm FS}}\left(\frac{E_0}{B_0} \right)=160\ {\rm MeV} \left(\frac{E_0}{B_0} \right),\end{equation} where $m_e$ is the electron mass,  $\alpha_{\rm FS}$ is the fine-structure constant, $c$ is the speed of light, $E_0$ is the typical electric field that accelerates particles, and $B_0$ is the typical magnetic field\footnote{Technically $B_0$ should be $B_{0,\bot}$ because only the magnetic field perpendicular to the particle's motion is responsible for synchrotron radiation. But unless there are special geometric constraints the perpendicular magnetic field is similar to the background magnetic field. In the case of Harris-like magnetic reconnection \citep{harris62} that is investigated in this paper, the motion of accelerating particles is primarily perpendicular to the reversing field. Unless a very strong guide field is present,  $B_0$ is therefore approximately equal to both the reversing field and the background field.} that is responsible for synchrotron radiation. In ideal plasmas like those involved in the simplest models of shock acceleration $E_0<B_0$, so synchrotron emission cannot be produced above  $\epsilon_{\rm bo, max}\equiv160$ MeV, the maximum synchrotron burnoff limit.

Yet there are many systems that appear to go above this limit.  Observations of gamma-ray flares in the Crab nebula  \citep{tavani_11,abdo_11,striani_11,buehler_12,mayer_13} present a clear case of emission beyond the synchrotron burnoff limit, with peak energies of 375 MeV with significant radiation up to 1 GeV \citep{buehler_12} and there is no evidence of ultrarelativistic motion there \citep{hester_02} to boost the photon energy. There are also other cases of high-energy emission which may involve synchrotron radiation.  The extended emission at energies of up to 300 GeV in GRBs \citep{ackermann_13} seems to require rest-frame photon energies above 160 MeV, although the emission may be inverse compton radiation \citep[e.g.,][]{wang_13}. TeV flares observed in AGN \citep[e.g.,][]{aharonian_07} seem to require relativistic motion within the jet to escape scattering in the emission region \citep{begelman_08,finke_08}, but they may also involve rest-frame synchrotron emission above the burnoff limit.
  
 In order to produce emission beyond 160 MeV, it is necessary to decouple particle acceleration and cooling so that the constraint $E_0<B_0$ no longer applies.   In the context of relativistic shocks, particles may accelerate in the background field but radiate in spatially varying or microturbulent fields near the shock in the synchrotron \citep{bykov_12,kumar_12,plotnikov_13,lemoine_13,lemoine_13b}  or jitter \citep{mao_11,teraki_13} regimes. But numerical simulations indicate that particle acceleration in relativistic shocks is unlikely to be efficient enough to produce synchrotron radiation beyond 160 MeV \citep{sironi_13}, and particles in these shocks are unlikely to emit in the jitter regime \citep{sironi_09}. 
 
  In contrast, magnetic reconnection naturally produces emission above the maximum synchrotron burnoff limit. In the diffusion region of magnetic reconnection, where much of the particle acceleration takes place, $B \ll B_0$  and $E_0>B$.  Significant cooling only occurs after the particle has left the acceleration region. Thus, particle acceleration and cooling are naturally decoupled in magnetic reconnection. Analytical studies of magnetic reconnection \citep{kirk_04, cont_07,uzdensky_11} and a test particle study in a fixed reconnection geometry \citep{cerutti_12a} indicate that particle acceleration through this mechanism is not limited by cooling. These studies suggest that particle acceleration in magnetic reconnection can produce synchrotron radiation beyond the maximum burnoff limit.

Particle-in-cell (PIC) simulations are the most common method for probing the acceleration of particles in fully dynamic magnetic reconnection.  These simulations, which focus on the pair-plasma case which is easier to simulate and is directly applicable to the Crab Nebula, have shown that relativistic reconnection can produce hard non-thermal tails of highly accelerated particles that are typically interpreted as power laws \citep{2001ApJ...562L..63Z,zenitani_05b,zenitani_07, zenitani_hesse_08b,jaroschek_04,jaroschek_08b,bessho_05,bessho_07,bessho_12,daughton_07,lyubarsky_liverts_08, liu_11, cerutti_12b, cerutti_13a, cerutti_14, werner_16, sironi_14,sironi_16,guo_14,guo_15,liu_15,yuan_16,lyutikov_16,kagan_16}.  In general, the physics of reconnection is similar in two and three dimensions \citep{sironi_14} unless the initial reconnection structure is inherently three-dimensional, justifying the use of two-dimensional simulations which reduce computational cost and allow larger-scale simulations to be carried out. Studies of the radiation produced by accelerated particles in reconnection that do not include the effect of cooling on particles \citep{cerutti_12b,kagan_16} indicate that radiation above the synchrotron burnoff limit should be highly beamed, and therefore highly variable, which is consistent with observations of fast variability in the Crab flares \citep{mayer_13}.

Several researchers have used PIC codes with a form of the radiation reaction force included \citep{hededal_05, tamburini_10} to investigate the effects of cooling\footnote{\citet{jaroschek_09} use a PIC simulation including another form of the radiation reaction force to investigate reconnection induced by the radiation reaction, but they do not focus on the effects of cooling on particle acceleration.} on particle acceleration in relativistic reconnection  \citep{cerutti_13a, cerutti_14,yuan_16}. They find that relativistic reconnection produces beamed particles and highly variable radiation above the synchrotron burnoff limit in both two \citep{cerutti_13a, yuan_16} and three  \citep{cerutti_14} dimensions. Relativistic reconnection also produces high-energy radiation above the synchrotron burnoff limit in the presence of a guide field \citet{cerutti_13a,cerutti_14} as long as it is weaker than the reversing field, and in reconnection produced by the collapse of a two-dimensional flux rope system \citep{yuan_16}. These simulations demonstrate that the production of radiation beyond the synchrotron burnoff limit is robust in relativistic magnetic reconnection. However, they are limited in dynamic range to $\le2$ decades in Lorentz factor $\gamma$. The peaks of the corresponding radiation spectra are located at or below the burnoff limit, with the high-energy tail reaching above the limit by less than a factor of 10 in frequency. Therefore, these simulations do not predict how far above the synchrotron burnoff limit radiation can be produced in relativistic reconnection.  We overcome this problem in our simulations using the new test particle comparison (TPC) method. 

In this paper we use two-dimensional PIC simulations including the radiation reaction force to reexamine the extent to which particles accelerated in relativistic reconnection can produce radiation above the synchrotron burnoff limit. In the absence of a large dynamic range in Lorentz factor, which is difficult to produce from an initial thermal spectrum even in the largest simulations \citep{sironi_14}, we instead choose to trace particle trajectories to separate the effects of cooling during acceleration (which limit the maximum radiation energy) from the effects of cooling after acceleration is complete (which do not limit it).   But tracing particles throughout the whole simulation imposes significant computational costs, and it is nontrivial to correct for the different sizes and characteristics of dynamically changing acceleration regions throughout the system, which are expected to influence the effects of cooling on particle trajectories and acceleration. 

Therefore, we carry out two simulations of relativistic reconnection in which the (uncooled) structural particles and fields that determine the evolution of the reconnection are identical, and compare the trajectories of test particles incident on a single typical reconnection region with and without cooling.   This TPC method allows us to isolate the effects of cooling on accelerating particles in a fully dynamic reconnection region, because its properties are the same in both simulations. We calculate a prescription for the effects of cooling based on a particle's uncooled trajectory in the reconnection region and the acceleration it experiences there.  We apply this prescription to analytically predict the effects of cooling on an uncooled power law energy spectrum produced in reconnection over a large dynamic range.   We qualitatively confirm our results over a limited dynamic range by comparing the particle spectra in these uncooled simulations with the spectrum in a simulation in which structural particles are cooled. 

Our paper is organised as follows. In Section \ref{sec:methods}, we discuss our simulation methods, including our treatment of the radiation reaction force and a detailed discussion of the TPC method. In Section \ref{sec:results}, we discuss the results,  and in Section \ref{sec:conclusions}, we discuss the conclusions of our research.

\section{Methodology} \label{sec:methods}
We use the particle-in-cell method to simulate reconnection. This method includes kinetic effects by discretizing the equations of electrodynamics (the mean-field Maxwell's Equations and the Lorentz Force Law), replacing groups of physical particles with larger macroparticles and tracking fields only on the vertices of a grid. We implement the PIC simulations using the {\tt Tristan-MP} code  \citep{spitkovsky_structure_2008}, which uses current filtering to greatly reduce noise even at relatively small macroparticle densities.  

We initialize all of our simulations with a simple Harris current sheet configuration that is susceptible to reconnection, described in Section \ref{sec:init}. Our first two simulations have identical evolution for the structural particles (which do not experience cooling) and fields. After nonlinear reconnection has developed, we inject test particles that duplicate structural particles flowing into a single reconnection region in the current sheet with (Simulation {\tt TC}) and without (Simulation {\tt TU}) applying cooling to those test particles.  Using this TPC method allows us to isolate the effects of cooling and predict its effects in large-scale simulations. To do this, we first find a prescription for the energy loss due to cooling for an accelerating particle as a function of the particle's trajectory. This is parameterized by 
\begin{equation}\xi\equiv\sqrt{\left \langle \frac{E B_0^2}{E_0 B^2}\right\rangle}.\end{equation}
Here the brackets indicate an average (described in Section \ref{sec:tpc}) over the particle trajectory in the acceleration region, $E$ and $B$ are the local electric and magnetic fields at each point in the trajectory, and $E_0$ and $B_0$ are the characteristic electric field of the reconnection region and the background magnetic field. 

We call $\xi$ the burrowing parameter, because it represents the extent to which the particle has burrowed into the current sheet and experienced a lower magnetic field than the background field. By finding the burrowing parameter as a function of the (uncooled) final energy of the test particles in the acceleration region, we can analytically predict the cooled particle energy spectrum from the uncooled spectrum, as explained in Section \ref{sec:tpc}. We also carry out a third simulation (Simulation {\tt AC}) in which cooling affects structural particles to evaluate the effects of cooling on the overall reconnection configuration during nonlinear reconnection and verify our analytical results. Table \ref{tab:overview} summarizes the properties of the three simulations.

  \begin{deluxetable}{ccccc}
  \tablecaption{Table of Simulations \label{tab:overview}}
\tablehead{\colhead{Run} &\colhead{Test Particles?} &\colhead{$\omegap t_{\rm 0,T}$ \tablenotemark{a}}&\colhead{Overall cooling?}&\colhead{Test Cooling?}}
\startdata 
{\tt TU}&Y&562.5&N&N\\
{\tt TC}&Y&562.5&N& Y\\
{\tt AC}&N&-&Y&N
\enddata
\tablenotetext{a}{ The time when test particles are injected in terms of the plasma frequency $\omegap$ defined in Section \ref{sec:params}.}
\end{deluxetable}

\subsection{Initial configuration} \label{sec:init}

The spatial domain is rectangular with $0\leq x<L_x$, $0\leq y<L_y$, and the boundary conditions are periodic in all directions.  The initial configuration contains two relativistic Harris current sheet equilibria \citep{harris62,kirk_03} without guide field at  $x=L_x/4$ and $x=3L_x/4$ with equal and opposite currents. Each current sheet has the magnetic field profile 

\begin{equation}
  {\mathbf B}=B_0  \tanh \left(\frac{x-x_0}{\Delta}\right) (\pm \hat{{\mathbf y}}),
\label{eq:harris_field}
\end{equation}
where $\Delta$ is the half-thickness of each current sheet and $x_0$ is its center. 

The density profile of the structural particles consists of a specially varying, drifting current population with maximum density $n_0$ centered at each current sheet, plus a background population of stationary particles of density $n_{\rm b}$:

\begin{equation}
n=n_0 \ {\rm sech}^2\left(\frac{x-x_0}{\Delta}\right) +n_{\rm b}. \label{eq:density_profile}
\end{equation}
Densities are defined including both species.

The resulting drift velocity of the positively and negatively charged current sheet particles is given by $\boldsymbol{\beta}_+=-\boldsymbol{\beta}_-=\pm B_0 /(4\pi n_0 q \Delta) (-\hat{{\mathbf z}})$., where $q$ is the charge of the macroparticle. For the magnetic field and drift velocity profiles, the sign in $\pm$ is positive for the vicinity of the  current sheet at $x=L_x/4$ and negative near the current sheet at $x=3 L_x/4$. The drifting and stationary populations of both species begin in relativistic Maxwellians at temperature $T_0=0.2 m c^2$.

\subsection{Reconnection Parameters}\label{sec:params}

The two most important parameters that determine the physics of reconnection in the absence of cooling are the magnetization in the background plasma
$\sigma_0$ 
\begin{equation}
\sigma_0\equiv \frac{B_0^2}{8\pi m n_{\rm b} c^2 h},
\label{eq:sigmadef}
\end{equation}
and the ratio $\Delta/\lambda_{\rm p}$ of the current sheet width to the plasma skin depth of particles in the center of the current sheet,

\begin{equation}
\lambda_{\rm p}=\sqrt{\frac{\langle\gamma\rangle m c^2}{4 \pi n_0 q^2 }}.
\label{eq:plasma}
\end{equation} 

Here $h=\langle\gamma\rangle +P/(m n c^2)$ is the average enthalpy of particles,  $\langle\gamma\rangle$ is the mean particle Lorentz factor in the current sheet, and $P$ is the particle pressure.

In order for fast relativistic reconnection to occur, one must have $\sigma_0\gg 1$ and $\Delta/\lambda_p\sim 1$. We choose parameters $\sigma_0=6.4$ and $\Delta/\lambda_{\rm p}=2$.  We choose a low value for $\sigma_0$ to probe the $\sigma_0 \lesssim 10$ regime for which the current sheet structure does not depend significantly on the initialization\footnote{Specifically, when $\sigma_0 \gg 10$ the current sheet becomes significantly wider and inflows become relativistic if the pressure gradient in the magnetic field is balanced by density gradients, but not if it is balanced by temperature gradients.} of the current sheet \citep{bessho_12,kagan_16}. $T_0=0.2 m c^2$ corresponds to $\langle\gamma\rangle=1.36$ and enthalpy $h=1.56$, so $\sigma_0=6.4$ gives $n_0/n_{\rm b}=50$.

The size of the simulations is $(L_x, L_y) =(1600 \lambdap, 1280\lambdap)$, large enough for nonlinear reconnection to occur while the boundary conditions do not affect the evolution. Test particles are injected in simulations {\tt TU} and {\tt TC} at $\omegap t=562.5$, and all simulations are run at least until $\omegap t=1000$, a time at which nonlinear reconnection begins to be affected by the boundary conditions. This is enough time to observe the effects of acceleration and cooling on the injected particles and evaluate the effects of cooling on the overall structure of the X-point regions.

To ensure sufficient resolution  to capture the physics of reconnection, we use a density of 8 macroparticles/cell/species throughout the plasma and a grid size $\Delta=\lambdap/8$. Tests of the code \citep{kagan_16} show that the physics of reconnection and the evolution of the current sheet is similar for macroparticle densities up to 50 macroparticles/cell/species and grid sizes as small as $\lambdap/20$.

\subsection{Radiative feedback}\label{sec:radfeedback}

We simulate the cooling of particles using the reduced Landau-Lifshitz radiation reaction force \citep{vranic_16}. { We express this force in dimensionless units with electric and magnetic fields normalized to a fiducial magnetic field $B_0$ and times normalized to $1/\omega_c$ (setting $\tau=\omega_c t$), where $\omega_c=qB_0/mc$ is the nonrelativistic gyrofrequency of particles in the field $B_0$}.  In terms of the normalized momentum $\mathbf{u}=\mathbf{p}/m c$  the radiation reaction force is given by 

\begin{align}
\left.\frac{d\mathbf{u}}{d\tau}\right |_{\rm rad} =&\zeta \bigg\{  -\left[ \left(\mathbf{E}+\frac{\mathbf{u}}{\gamma}\times \mathbf{B}\right)^2-\frac{(\mathbf{E} \cdot \mathbf{u})^2}{\gamma^2}\right]\gamma \mathbf{u} \label{eq:reactionforce}\\
&+ \mathbf{E}\times \mathbf{B}+\frac{1}{\gamma} \mathbf{B} \times (\mathbf{B}\times \mathbf{u})+\frac{1}{\gamma}\mathbf{E}(\mathbf{u} \cdot \mathbf{E})\bigg\},\nonumber
\end{align}
where $\zeta \equiv 2 q^3 B_0 /(3 m^2 c^4)$.

The terms on the first line represent the \citet{hededal_05} radiation reaction force used by many other authors \citep{sironi_09, cerutti_13a, cerutti_14,yuan_16}. These terms typically dominate the radiation reaction force, but we include the additional reduced Landau-Lifshitz terms on the second line to increase the accuracy of the simulation without significant computational cost. We have tested the cooling code extensively in the synchrotron, bremstrahlung, and curvature radiation regimes, finding good agreement with analytical expectations. We do not include the effects of inverse Compton cooling, because in a realistic case in which electrons emit close to the burnoff limit, the radiation is deep in the Klein-Nishina regime.

Because Maxwell's Equations and the Lorentz force law are linear in the fields $\mathbf{E}$, $\mathbf{B}$,   { charge density $\rho$, and current density $\mathbf{J}$ in the absence of cooling, $B_0$ and $n_b$ can be scaled out of the equations of motion using a similar procedure to that in \citet{yuan_16}.  We first normalize the electric and magnetic fields to $B_0$. We then scale times to $1/\omega_c$ (setting $\tau=\omega_c t$) and distances to $c/\omega_c$. We normalize the phase space density to the background number density $n_b $ and then express it in terms of dimensionless positions, momenta, and times as $f(\mathbf{r} \omega_c/c,\mathbf{u},\tau)$ where $\mathbf{r}=(x,y,z)$. Our choice of normalization for the phase space density determines the normalization of its moments:  the number density is normalized to $n_b$, the charge density to $q n_{\rm b}$ and the current density to $q n_{\rm b} c$.    We find that the resulting dimensionless Vlasov-Maxwell equations for the normalized fields in the absence of the radiation reaction force are}

\begin{align}
\frac{\partial f }{\partial \tau}+(\mathbf{u} \cdot \mathbf{\nabla}) f &+\frac{d\mathbf{u} }{d\tau}\cdot \mathbf{\nabla_u}  f=0,\\
\frac{d \mathbf{u} }{d\tau}&= \frac{q}{|q|} \left(\mathbf{E} +\frac{\mathbf{u}}{\gamma}\times \mathbf{B}\right),
\end{align}

\begin{align}\mathbf{\nabla \cdot E} &= \frac{{\rho}}{2\sigma_0 h}, \\
\mathbf{\nabla} \cdot \mathbf{B}&= 0,\\
\mathbf{\nabla} \times \mathbf{E}& = -\frac{\partial \mathbf{B}}{\partial \tau}, \\
\mathbf{\nabla} \times \mathbf{B}&=\frac{\partial \mathbf{E}}{\partial \tau} + \frac{\mathbf{J}}{2\sigma_0 h},
\end{align}
{ where the differential operators are defined as $\mathbf{\nabla_u}\equiv(\partial/\partial u_x,\partial/\partial u_y,\partial/\partial u_z)$ and $\mathbf{\nabla}\equiv(\partial/\partial (x \omega_c/c),\partial/\partial(y \omega_c/c),\partial/\partial (z \omega_c/c))$. 

The only parameter that appears in the equations is $2\sigma_0 h=B_0^2/(4\pi n_{\rm b} m c^2)$. The fiducial magnetic field $B_0$ and background number density $n_{\rm b}$ thus appear only in combination, and either may be varied without changing the form of the simulation as long as a corresponding adjustment is made to the other. The setup of the initial conditions also depends on the normalized current sheet width $\Delta/\lambdap$ and the initial normalized temperature $T_0/(m c^2)$. Combined with the size of the simulation, these parameters completely characterise our simulations in the absence of radiative feedback.}

{ With the radiation reaction force added,  $\zeta$ is a new dimensionless parameter which sets the limiting Lorentz factor that corresponds to synchrotron emission at the burnoff limit. The value of $\zeta$ also constitutes the choice of a physical value of the fiducial magnetic field $B_0$ for a given species of physical particle, because the choice of a particle sets $\zeta/B_0=2q^3/(3m^2c^4)$.}

We derive the limiting Lorentz factor as follows: a relativistic particle accelerated in the typical electric field $E_0$ has an energy gain per unit time of approximately. 

\begin{equation}
m c^2\left(\frac{d\gamma}{dt}\right)_{\rm accel}=q E_0 c,\end{equation}
while the same particle radiating in the typical background magnetic field $B_0$ loses energy at a rate
\begin{equation} m c^2\left(\frac{d\gamma}{dt}\right)_{\rm rad}=-\frac{2q^4 B_0^2 \gamma^2}{3 m^2 c^3}. \label{eq:coolingfirst}\end{equation}

The critical Lorentz factor for which the acceleration and cooling are equal is then given by
\begin{equation}\gamma_{\rm bo,0}=\sqrt{\frac{3 m^2 c^4 E_0} {2 q^3 B_0^2}}. \label{eq:lorentzbo}\end{equation}

The synchrotron peak of particles with this Lorentz factor is at the burnoff limit $\epsilon_{\rm bo,0}$, so we call $\gamma_{\rm bo,0}$ the fiducial burnoff limit for particle energy. We can express $\zeta$ in terms of the limiting Lorentz factor as

\begin{equation}
\zeta=\frac{ E_0}{ B_0 \gamma_{\rm bo,0}^2}   , \label{eq:zeta}
\end{equation}

This allows us to choose a value for $\zeta$ (and thus, $B_0$) based on the ratio of the reconnection electric field to the background field $E_0/B_0$ and the desired value for $\gamma_{\rm bo,0}$. In simulations {\tt TC} and {\tt AC}, we have chosen $\gamma_{\rm bo,0}=12.6\sqrt{E_0/B_0}$.  We find similar qualitative results in both simulations for a larger value of $\gamma_{\rm bo,0}=32.6\sqrt{E_0/B_0}$. 

We now check the validity of our expression (\ref{eq:reactionforce}) for the radiation reaction force. In order for its derivation to be valid, it is necessary that particles are not strongly perturbed by the magnetic field on the light-crossing time of their classical electromagnetic radii. This condition, stated mathematically, is
\begin{equation}
B_p\ll\frac{m^2c^4}{q^3},
\end{equation}
where $B_p$ is the magnetic field measured in the rest frame of the particle. 
Using Equation (\ref{eq:lorentzbo}), we find  the radiation reaction force equation (7) is valid if throughout the simulation all particles have
\begin{equation}
B_p \ll B_{c}\equiv B_0 \gamma_{\rm bo, 0}^2\frac{B_0}{E_0}, \label{eq:bpcrit}
\end{equation}
where $B_{c}$ is the critical magnetic field at which Equation (\ref{eq:reactionforce}) becomes invalid. For $\gamma_{\rm bo,0}=12.6\sqrt{B_0/E_0}$, we find that $B_c = 159 B_0$.

We now verify that this condition is met in our simulations. For all of our traced particles in Simulation {\tt TC}, we use Lorentz transformations to calculate $B_p$ at the particle location from the local value of $B$ in the simulation frame. Over the complete particle trajectories, we find that in all cases $B_p<20.9 B_0$, which is factor of $\sim8$ below $B_c$. Because our test particle analysis is based only on the particle trajectories during acceleration, this is actually an overestimate of the relevant values of $B_p$. For all particles during the acceleration phase, we find that at all times $B_p<8.7B_0$, which is a factor of $\sim18$ below $B_c$. Thus, our treatment of the radiation reaction force is accurate in our test particle analysis. In realistic physical situations where $\gamma_{\rm bo,0}\ggg1$, $B_p$ will be even smaller compared to $B_c$ for particles near the burnoff limit because $B_c/B_0 \propto \gamma_{\rm bo,0}^2$. While quantum effects are expected to set in below $B_c$ (at $\alpha_{FS} B_c\approx B_c/137$ for electrons), they are unimportant in this realistic case. Therefore, including quantum effects in our simulations would be incorrect and we do not do so.

 We also calculate $B_p$ for all particles in Simulation {\tt AC} at time $\omegap t=562.5$. Due to the presence of regions of very strong magnetic field of up to $10 B_0$ at the edges of islands, a significant number of particles do experience fields approaching $B_c$.  Quantitatively, approximately $23\%$ of the accelerated particles with $\gamma>10$ have $B_p>B_c/4$. However, particles in X-point regions in this simulation always have $B_p<B_c/4$, so the physics of particle acceleration and the high-energy part of the particle energy spectrum are unaffected by inaccuracies arising from the use of the classical radiation reaction force. We then compare of Simulation {\tt AC} with a simulation that is identical except for larger $\gamma_{\rm bo,0}\sqrt{B_0/E_0}=32.6$ to verify that cooling does not greatly change the current sheet structure. We find that the small qualitative effects resulting from cooling found in Simulation {\tt AC} are verified in this additional simulation, and they are quantitatively even smaller as expected due to the weaker cooling. The reason that the inaccuracies have little effect is that they merely affect the details of how particles lose energy in islands after the end of particle acceleration. Because the purpose of Simulation {\tt AC} is merely to test that particles can be accelerated beyond the burnoff limit and verify that cooling does not strongly affect the current sheet structure, we reserve more rigorous fully cooled simulations for future work.
 
 

Background particles in Simulation {\tt AC} are mildly relativistic and can cool slowly over the time of the simulation. This can distort the reconnection configuration in a way that would not occur in a realistic case in which the burnoff limit corresponds to a much higher Lorentz factor $\gamma_{\rm bo,0}\ggg 1$. Therefore, we turn off cooling in Simulation {\tt AC} for particles that are below $\gamma=5/3$ to prevent background cooling from affecting the current sheet structure.
\subsection{The Test Particle Comparison (TPC) method}\label{sec:tpc}\begin{figure*}
\begin{center}
\includegraphics[width = 0.8\textwidth]{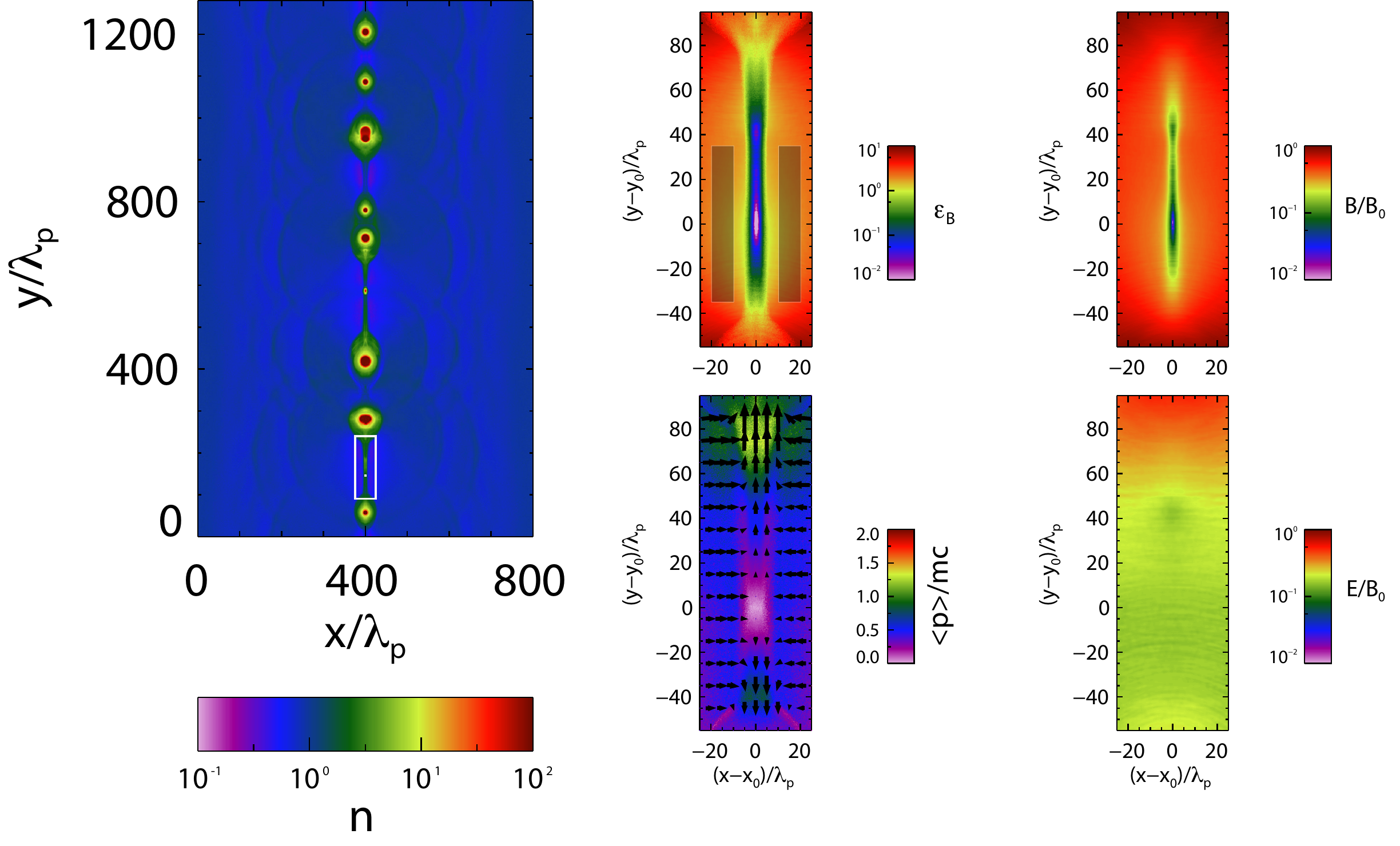}
\end{center}
\caption{Properties of the current sheet at the beginning of particle tracing at $\omegap t=562.5$.
(left) The normalized density $n$ throughout the (half) simulation box. The white outline shows the asymmetric X-point region for which we carry out particle tracing, while the white square shows the center of the X-point at $(x_0=400 \lambdap, y_0=145 \lambdap)$. (right) Characteristics of the X-point region. The magnetic energy to total particle rest mass ratio  $\epsilon_{\rm B}=B^2/8 \pi n m c^2$ (top left),  the bulk momentum of the plasma $\langle p \rangle/mc$ (bottom left) with corresponding directional arrows in the selected X-point region, the normalized magnetic field $B/B_0$ (top right), and the normalized electric field $E/B_0$  The shaded boxes in the top left panel show where traced test particles are injected.
\label{fig:overall}}
\end{figure*}
The TPC method compares initially identical test particles entering a reconnection region with (Simulation {\tt TC}) and without (Simulation {\tt TU}) cooling to calculate the effect of cooling on a chosen uncooled spectrum over a large dynamic range. It assumes that the trajectories of these test particles, and the values of $\xi$ (defined below) that parameterize those trajectories, do not depend significantly on whether cooling is present. Using this assumption, we can predict the maximum Lorentz factor reached by a particle in simulation particle as a function of its trajectory and final Lorentz factor in the uncooled simulation.

The local limit on particle acceleration resulting from cooling at a given point is given by Equation (\ref{eq:lorentzbo}), with the local values of the fields replacing $E_0$ and $B_0$.  The resulting local burnoff limit is given by $\xi_s\gamma_{\rm bo,0} $, where $\xi_s\equiv \sqrt{ E B_0^2/(E_0 B^2)}$. We hypothesize that the global limiting Lorentz factor for the particle is $\xi\gamma_{\rm bo,0}$, where $\xi$, called the burrowing parameter, is an average of $\xi_s$ over the particle's trajectory. Because we are interested in the values of $\xi_s$ at locations where the particle radiates, we weight the average by the synchrotron power of the particle. In a simple approximation, synchrotron power is proportional to $B^2\gamma^2$, so we define 

\begin{equation}
\xi\equiv\frac{\int{\xi_s B^2\gamma^2} dt}{\int{B^2 \gamma^2 dt}}.
\end{equation}

We then predict that the final Lorentz factor $\bar{\gamma}_{\scriptscriptstyle \mathrm{f}}$ reached by a particle in the cooled simulation is given by
\begin{equation}
\bar{\gamma}_{\scriptscriptstyle \mathrm{f}} ={\rm min}(\gamma_{\scriptscriptstyle \mathrm{f}}, \xi\gamma_{\rm bo,0}). \label{eq:prescription}
\end{equation}

The f subscript is present because we are referring to the final Lorentz factor measured for the particle at the end of acceleration in each simulations. Because both $\xi$ and $\gamma_{\scriptscriptstyle \mathrm{f}}$ on the right side of the equation are calculated in the uncooled Simulation {\tt TU}, this equation can be used to predict the effects of cooling in a simulation with a given value of  $\gamma_{\rm bo,0}$ without actually carrying out that simulation.

 The final step of the TPC method is the calculation of the distribution function for $\xi$ at any given value of $\gamma_{{\textrm f}}$ using the measured trajectories. For any given uncooled particle energy spectrum, we can use this distribution function and the prescription (\ref{eq:prescription}) to derive the cooled particle energy spectrum. In Section 3.5, we describe in detail how to derive the cooled spectrum corresponding to an uncooled power law energy spectrum and the distribution for $\xi$ at each $\gamma_{{\scriptscriptstyle \mathrm{f}}}$ that we find in our simulations. 
  \section{Results} \label{sec:results}
 
We begin in Section \ref{sec:structure} with a description of the overall structure of the current sheet in Simulation {\tt TU} as well as the characteristics of the X-point region we focus on in our analysis. Then we describe our analysis of cooling during particle acceleration using the TPC method. This consists of a study of the effects of cooling on typical particle trajectories (Section \ref{sec:trajectories}), a model of the effects of cooling on particle acceleration (Section \ref{sec:model}), and a calculation of the relationship between particle trajectories and particle acceleration in the absence of cooling (Section \ref{sec:relationship}). In Section \ref{sec:powerlaw}, we use this analysis to analytically compute the effect of cooling on a power law particle energy distribution. Finally, in Section \ref{sec:comparison}, we compare the current sheet structure and particle energy spectra in Simulations {\tt TU} without cooling and {\tt AC} with cooling to test the effects of cooling in a full simulation.

\subsection{Structure of the X-point region}\label{sec:structure}

The left part of Figure \ref{fig:overall} shows the structure of simulation {\tt TU} (which is the same as that of Simulation {\tt TC}) during the nonlinear phase of magnetic reconnection  at time $\omegap t=562.5$ when particle tracing begins. As in other simulations of collisionless reconnection, the current sheet has broken up into many X-point regions (where reconnection takes place) and high-density magnetic islands (where outflows from the X-point regions meet).  The right part of the figure shows an inset of the asymmetric X-point region into which test particles are injected in both simulations. The size of the X-point region is small compared to the overall box size, ensuring that its dynamics and our test particle results are unaffected by the boundary conditions.  

\begin{figure*}
\begin{center}
\includegraphics[width = 0.8\textwidth]{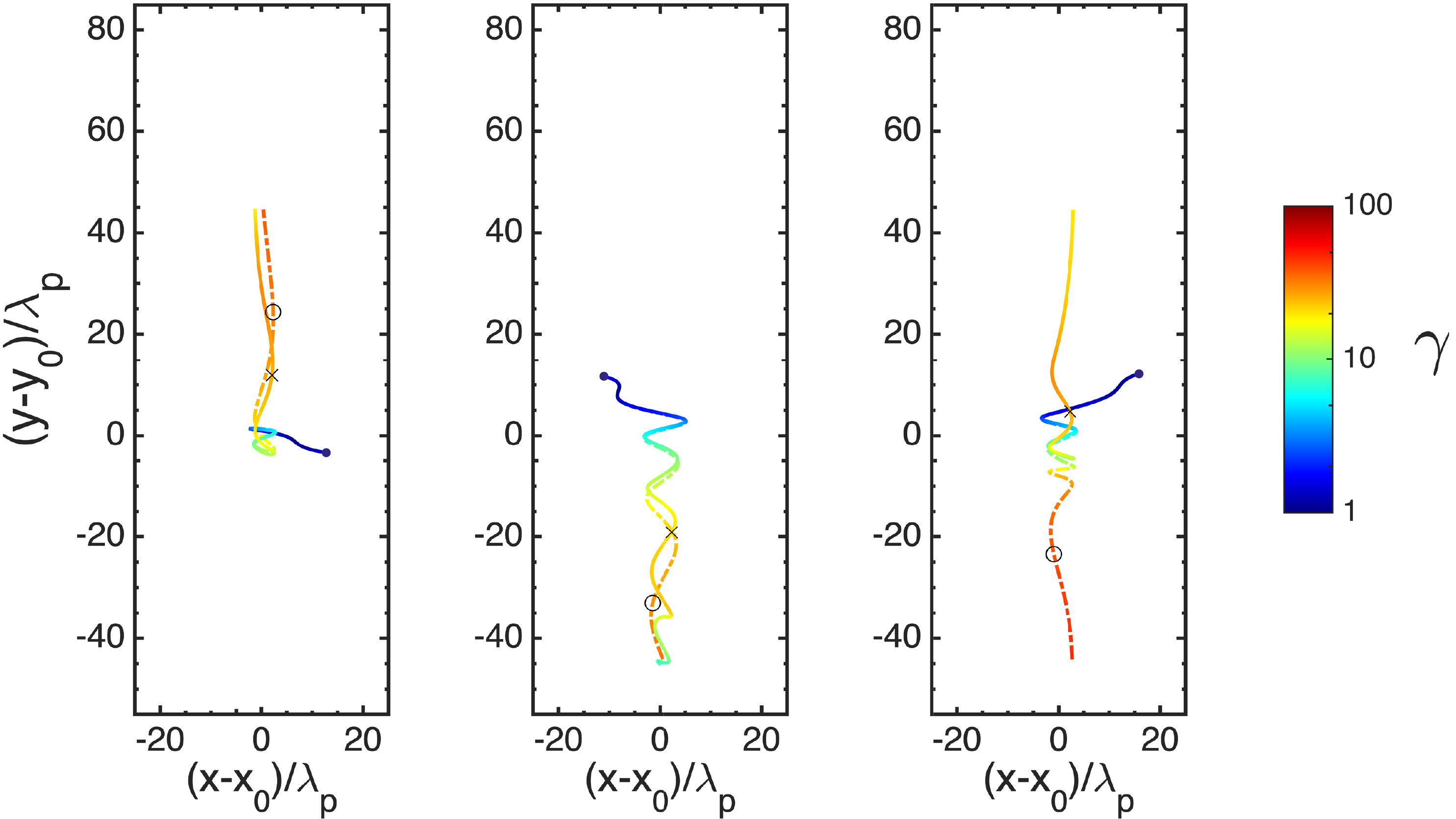}
\end{center}
\caption{Trajectories within the reconnection region shown in the right half of Figure 1 for three particles in the cooled Simulation {\tt TC} (solid line) and uncooled Simulation {\tt TU} (dot-dashed line).  Colors indicate the Lorentz factor at each point in the particles' trajectories. The circle (x) marks indicate the location where acceleration of the cooled (uncooled) particle ends and the final Lorentz factor is calculated using the procedure discussed in Section \ref{sec:trajectories}. The large dots indicate where the particle trajectories begin, which is the same location in both simulations.}
\label{fig:trajectories}
\end{figure*}

The structure of the X-point region is best indicated by comparing the plot of $\epsilon_{\rm B}=B^2/8 \pi n m c^2$ (top left)  with that of $B/B_0$  (top right). The narrow reconnection region at , $-45<(y-y_0)/\lambdap<45$ in the central part of the X-point region is identical in both plots, indicating that the magnetic field is far below the background as expected in reconnection regions.  In contrast, in the outflow regions of the figure with $|y-y_0|/\lambdap>45$ the magnetic field is similar to that in the background but the density is enhanced, producing a lower $\epsilon_{\rm B}$ that is consistent with expectations for outflow regions where particles are being ejected into the magnetic islands. The lower left panel shows the momentum structure of the reconnection region. This clearly indicates that the center of the X-point is indeed at $x=x_0,\ y=y_0$. The maximum inflow momentum we find for this X-point region is $\langle p \rangle/mc \sim 0.4$, while typical outflow bulk momenta are  $\langle p\rangle/mc\sim1.4$. This is similar to the values found in \citet{kagan_16} for $\sigma_0=4$.

In order to calibrate our analysis of the test particle trajectories, we must find the value of $E_0/B_0$ in the X-point region. The lower right panel of the figure shows that the electric field is approximately uniform with $E/B_0\approx 0.174$ throughout the central reconnection region; therefore, we set $E_0/B_0=0.174$ for this region. Since we chose $\gamma_{\rm bo,0}=12.6\sqrt{E_0/B_0}$ for our simulations, the fiducial burnoff limit for particles in this X-point region is $\gamma_{\rm bo,0}=5.45$. $E_0/B_0$ is also a measurement of the reconnection rate, because in a steady state reconnection equilibrium $E_0=(v_{\rm in}/c)B_0$. We also consider the results of two other methods of measuring the reconnection rate. One involves calculating the ratio of the inflow velocity to the Alfv{\'e}n velocity in the inflow region (which is close to $c$ for relativistic plasmas); with this definition, the typical reconnection rate shown in the lower left panel of Figure 1 is $v_{\rm in}/c \approx0.4$ . However, this is a local measure that overestimates the overall rate of energy conversion. Another, global, method involves comparing the total kinetic energy gain to the maximum possible gain \citep{kagan_16}, with the equation

\begin{equation}
r_{\rm rec}=\frac {d \mathcal{E}_{\rm K}}{dt}\frac{L_x}{4 c \mathcal{E}_{\rm B,0}},
\end{equation}
where $\mathcal{E}_{\rm K}$ is the total kinetic energy and $\mathcal{E}_{\rm B,0}$ is the total initial magnetic energy. With this definition, we find $r_{\rm rec}=0.11$, which is consistent with (but somewhat smaller than) the reconnection rate calculated from $E_0/B_0$. This indicates that the energy conversion rate in our chosen X-point is representative of that in the overall simulation.

\subsection{Comparison of cooled and uncooled trajectories.}\label{sec:trajectories}
We select a sample of $\sim26000$ particles from the test particles the were initiated in the shaded regions in Figure \ref{fig:overall} and trace their trajectories within our chosen reconnection region in the cooled Simulation {\tt TC} and uncooled Simulation {\tt TU}   The particles selected are approximately uniformly distributed in $\log \gamma$ at $\omegap t=1125$ well after all of the particles have left the X-point region. This enables us to calculate the properties of particles that have undergone differing amounts of acceleration with reasonable number statistics. We choose to calculate the properties of the electrons in the rest of this work, but the properties of the positrons are similar. 

We calculate the final Lorentz factors for particles in the uncooled ($\gamma_{\scriptscriptstyle \mathrm{f}}$) and cooled ($\bar{\gamma}_{\scriptscriptstyle \mathrm{f}}$) simulations as follows. To ensure that we calculate the final Lorentz factor for the main acceleration episode for each particle and properly estimate the values of the fields during this episode, we restrict our analysis of trajectories to particles in the central X-point region at $-45<(y-y_0)/\lambdap<45$, where they accelerate in a region of low magnetic field. This restriction does not have a strong effect on our results, because particles do not spend much time or undergo a great deal of acceleration in the region $45<(y-y_0)/\lambdap<80$.  We then estimate the Lorentz factor at the end of the acceleration episode by calculating the location where it reaches $85\%$ of its maximum value. This allows us to avoid including locations where the particle Lorentz factor plateaus, resulting in an underestimation of the average electric field acceleration. Because particles in the uncooled simulation may undergo multiple acceleration episodes, we add an additional stipulation: if the rate of energy gain by a particle falls below a fraction $1/e$ of the fiducial gain given by $qE_0 c$ and the particle has reached $60\%$ of its maximum value in the X-point, we estimate that acceleration has ended at that location. 

 Figure \ref{fig:trajectories} compares the trajectories within the chosen reconnection region of three particles cooled Simulation {\tt TC} and uncooled Simulation {\tt TU}; all of the particles chosen were significantly accelerated in the current sheet with $\gamma_{\scriptscriptstyle \mathrm{f}},\bar{\gamma}_{\scriptscriptstyle \mathrm{f}}>15$ and significantly cooled before leaving the current sheet in the cooled simulations. The fact that a significant number of particles can reach such high energies is an indication that radiation does not directly prevent acceleration beyond the fiducial burnoff limit in magnetic reconnection. The figure shows that all of the particles undergo Speiser orbits: they are accelerated by the reconnection electric field while oscillating across the current sheet in the $\pm x$ direction. The oscillation amplitude decreases with time as the particles are accelerated in accordance with momentum conservation constraints \citep{uzdensky_11}.
 
  The first and second panels show that the trajectories in the cooled and uncooled simulations are typically similar at the beginning of acceleration and while the particles are at the center of the X-point region, but sometimes differ after most acceleration is complete. Particles' cooled and uncooled trajectories typically differ most in the phase of their oscillation across the current sheet, but cooled particles also tend to leave the X-point slightly earlier than the uncooled particles, probably because cooling has reduced their inertia against deflection out of the X-point region.  The third panel shows the relatively rare case in which the cooled and uncooled particles exit in different directions. Such particles are only around $0.3\%$ of all particles, but they are a somewhat larger proportion ($1.1\%$) of the highly accelerated particles with $\gamma_{\scriptscriptstyle \mathrm{f}},\bar{\gamma}_{\scriptscriptstyle \mathrm{f}}>15$. Because few particles have trajectories that are strongly modified by cooling, it should be possible to use the properties of the particles in the uncooled simulation to predict the acceleration and cooling of the particles in the cooled simulation.

 \subsection{Model of cooling during particle acceleration} \label{sec:model}

The TPC method can be used to predict the final energy reached by cooled particles from the corresponding uncooled particle energies and trajectories. It relies on the assumption that $\xi$ is the same in both the cooled and uncooled simulations. We verify that $\xi$ is indeed nearly the same in the cooled Simulation {\tt TC} and the uncooled Simulation {\tt TU}. This result is in agreement with the similarity of the particle trajectories in the two simulations, discussed in Section \ref{sec:trajectories}.  

\begin{figure}[H]
\begin{center}
\includegraphics[width = 0.45\textwidth]{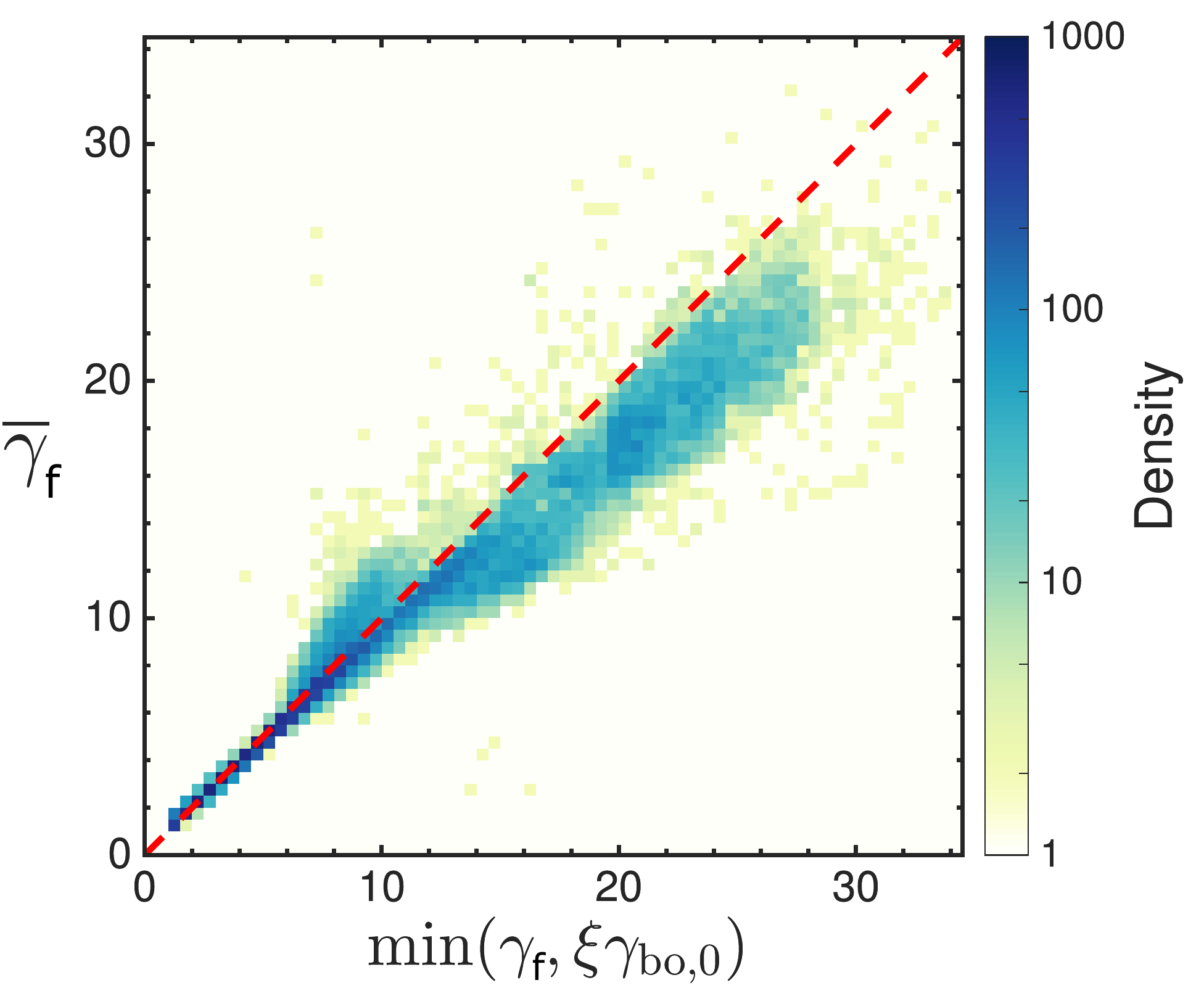}
\end{center}
\caption{The relationship between the predicted cooled Lorentz factor ${\rm min}(\gamma_{\scriptscriptstyle \mathrm{f}}, \xi\gamma_{\rm bo,0})$ and the actual cooled Lorentz factor $\bar{\gamma}_{\scriptscriptstyle \mathrm{f}}$ at the end of acceleration. The dashed line indicates where the two are equal, showing that the prediction is fairly accurate.
\label{fig:predcool}}
\end{figure}

Figure \ref{fig:predcool} shows a two-dimensional histogram of the relationship between the predicted and actual Lorentz final Lorentz factors in the cooled simulation. It shows that the model indeed predicts the cooled particle Lorentz factor of most particles very well, although it tends to slightly underestimate (by up to $20\%$) the amount of cooling for high-energy particles. Even restricting the calculation to particles that have undergone significant cooling so that $\xi \gamma_{\rm bo}<\gamma_{\scriptscriptstyle \mathrm{f}}$ produces a similar relationship. The great accuracy of the model indicates that our prescription (\ref{eq:prescription}) captures the physics of radiative cooling of particles within an X-point region. 
\subsection{Relationship between $\xi$ and $\gamma_{\scriptscriptstyle \mathrm{f}}$} \label{sec:relationship}

\begin{figure}[h]
\begin{center}
\includegraphics[width = 0.45\textwidth]{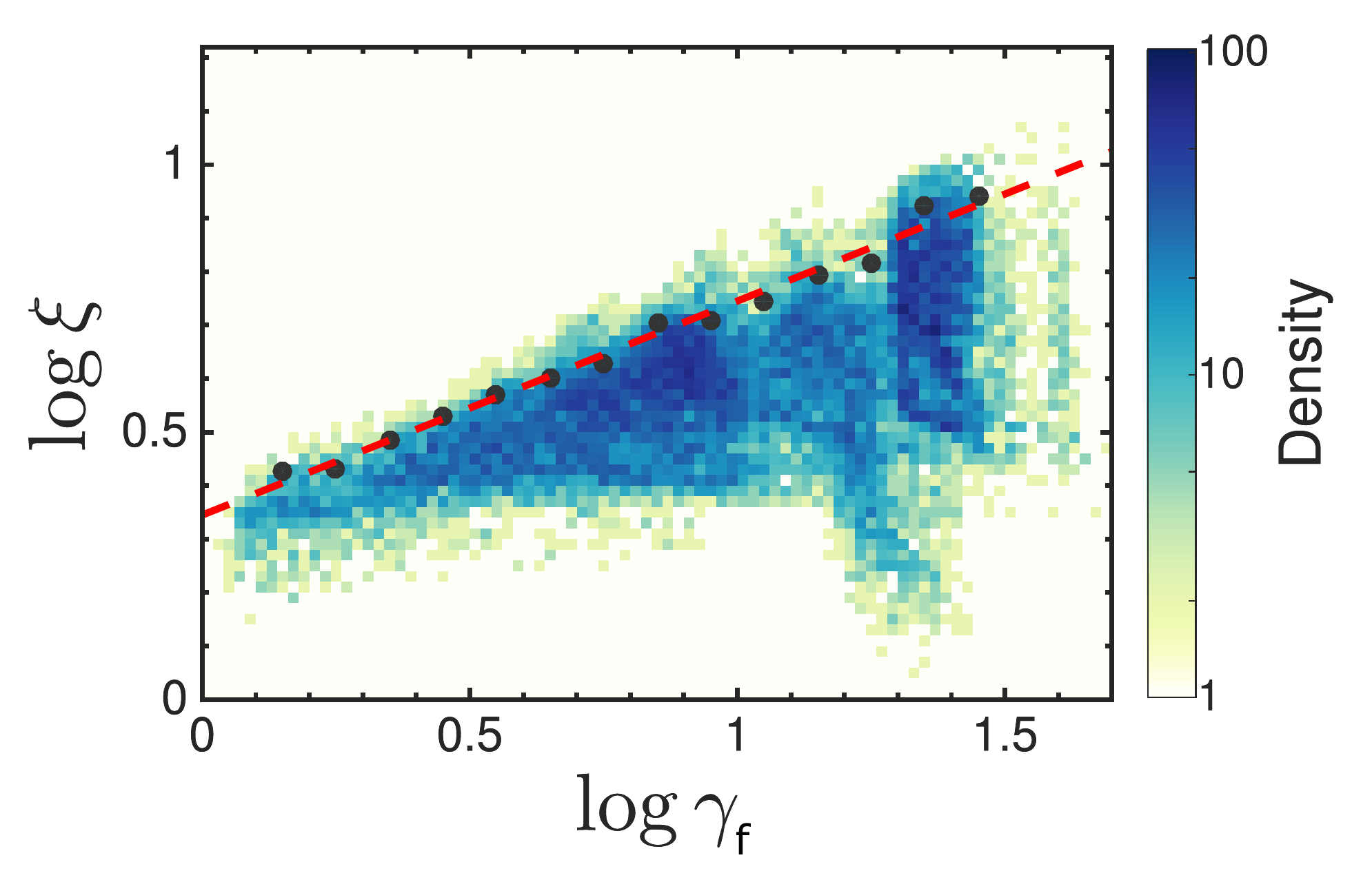}
\end{center}
\caption{The relationship between $\xi$ and uncooled Lorentz factor ${\gamma}_{\scriptscriptstyle \mathrm{f}}$. The distribution of $\xi$ is uniform at each value of $\gamma_{\scriptscriptstyle \mathrm{f}}$ for $2<\xi<\xi_{\rm max}$. The circles show the value of $\xi_{\rm max}$ for each interval of 0.1 in $\log\gamma_{\scriptscriptstyle \mathrm{f}}$. No circles are shown for $\log \gamma_{\scriptscriptstyle \mathrm{f}}=0-0.1$ or $\log \gamma_{\scriptscriptstyle \mathrm{f}}>1.5$ due to insufficient statistics. The red line is a fit of $\log \xi_{\rm max}$ vs. $\log \gamma_{\scriptscriptstyle \mathrm{f}}$ with a slope of $0.40$.
\label{fig:rhogamma}}
\end{figure}

Figure \ref{fig:rhogamma} shows the relationship between $\xi$ and the uncooled final Lorentz factor $ \gamma$. It shows that the minimum value of $\xi$ for all values of $\gamma_{\scriptscriptstyle \mathrm{f}}$ is $\xi_{\rm min}\sim 2$, and the probability distribution for $\xi(\gamma)$ is uniform up to a maximum value of $\xi$ that increases with $\gamma_{\scriptscriptstyle \mathrm{f}}$. A small proportion ($\sim 3 \%$) of the particles with $\gamma_{\scriptscriptstyle \mathrm{f}}>15$  have $\xi<\xi_{\rm min}$; they have undergone multiple acceleration episodes and therefore spent significant time in regions of low $\xi$ while still undergoing acceleration.  We calculate $\xi_{\rm max}(\gamma_{\scriptscriptstyle \mathrm{f}})$ by finding the value of $\xi$ where the distribution falls to $1/e$ times the peak at each interval of  0.1 in $\log \gamma_{\scriptscriptstyle \mathrm{f}}$.   A power law fit for $\xi_{\rm max} \propto \gamma_{\scriptscriptstyle \mathrm{f}}^{\alpha}$ yields $\alpha=0.40$. 

This distribution can be compared with analytical models of Speiser orbits, which fall into two classes. In the first, the particle's oscillation amplitude is always larger than the current sheet width $\Delta$, so that $\alpha=0$. In the other case, in which the amplitude of oscillation is comparable to or smaller than the current sheet width and $ B \sim B_0 \Delta/x_{\rm max}$, where $x_{\rm max}$ is the amplitude of oscillation for the particle. We can calculate the value of $\xi$ by finding $x_{\rm max}(\gamma)$ during the particle trajectory analytically and using the relation

\begin{equation} \langle B^2\rangle=B_0^2 \int^{x_{\rm max}/\Delta}_{-x_{\rm max}/\Delta} x^2 dx \sim B_0^2 \left(\frac{x_{\rm max}}{\Delta}\right)^3, \label{eq:speisbsq}\end{equation}
which for $E\approx E_0$ gives $\xi \sim (x_{\rm max}/\Delta)^{-3/2}$.

We calculate $x_{\rm max}(\gamma)$ using modified version of the analytical arguments of \citet{uzdensky_11}. We assume that the angle $\theta$ at which the particle exits the current sheet center during each oscillation is small and the particle's velocity in the $z$ direction is close to $c$, so that $dz/dt =c$ and $dy/dz = \theta$. Defining $\bar{z}=z \omega_c /c$ and  $\bar{x}=x \omega_c /c$, the equations of motion of the particle are then
 
 \begin{equation}
 \frac{d(\gamma v_x)}{d \bar{z}}=-\frac{x_{\rm max}}{\Delta}. \label{eq:speismot}
 \end{equation} 
 \begin{equation}
\frac{d\gamma}{d{\bar z}}=r_{\rm rec}.
 \end{equation}
 
 Using an analagous argument to that in \citet{uzdensky_11}, we find as they do that $\theta \propto \gamma^{-2/3} $. But we find a different expression for $x_{\rm max}$:
 \begin{equation}
 x_{\rm max} \sim  \gamma \theta^2 \frac{\Delta}{x_{\rm max}}.
 \end{equation}
 
Combining these expressions with Equation (\ref{eq:speisbsq}), we obtain $x_{\rm max} \propto \gamma^{-1/6}$ and $\alpha=0.25$.
 
 In Figure \ref{fig:rhogamma}, we see that there are always particles with $\xi\approx 1$ for all $\gamma_{\scriptscriptstyle \mathrm{f}}$; these particles typically never burrow into the current sheet and correspond to our first regime. The particles that do burrow into the current sheet have $\xi_{\rm max} \propto \gamma_{\scriptscriptstyle \mathrm{f}}^{\alpha}$ with $\alpha=0.4$, which is slightly larger than the analytical value of $\alpha=0.25$ in the second regime. We note that in the test particle simulations of \citet{cerutti_12a}, the oscillation amplitude declined somewhat faster than analytical expectations in the regime where radiation is negligible. That result is consistent with what we find here.

   \subsection{Effects of cooling on a power law particle distribution} \label{sec:powerlaw}

We now extrapolate the results of our simulation using analytical techniques to calculate the effects of cooling on particle energy spectra over many decades in dynamic range. Because we are calculating a steady-state distribution, we omit the subscript f for the cooled and uncooled Lorentz factors.

We assume that the  uncooled distribution of the particles is a pure power law of the form $N(\gamma)=  \gamma^{-p}$, where
$\gamma$ here is the uncooled Lorentz factor of a particle normalised to the minimum Lorentz factor $\gamma_{\rm min}$ of the power law. We assume that $\xi$ has a characteristic value $\xi_{\rm max}$ that increases with $\gamma$ as a power law, so $\xi_{\rm max}(\gamma)\propto \gamma^{\alpha}$, as discussed in the previous section. 
The measured distribution of $\xi(\gamma)$ in Figure \ref{fig:rhogamma}  indicates that the conditional distribution of $\xi$ at a given value of $\gamma$ is approximately uniform over $\xi_{\rm min}<\xi<\xi_{\rm max}$, and 0 elsewhere, so that  $\xi_{\rm max}(1) =\xi_{\rm min}$. The actual value of $\xi_{\rm min}\approx 2$ in the figure, but we assume $\xi_{\rm min}=1$ in our analysis for simplicity. This does not significantly affect\footnote{For $\xi_{\rm min}=k$, the locations of the breaks in the particle energy spectrum in Equation (\ref{eq:fulldist}) are scaled by a factor of $k$ and $k^{1/(1-\alpha)}$, but the functional forms do not change. } our results.  The resulting joint probability distribution $N(\gamma,\xi)$ for $\gamma>1$ is 

\begin{equation}
N(\gamma, \xi)=(p-1) \frac{\gamma^{-p}}{\gamma^{\alpha}-1}.   \ \ \ \ 1<\xi<\gamma^{\alpha}
\end{equation}
  We also set $N(1,\xi)=\delta(\xi-1)$ where $\delta$ here refers to the Delta function, so that $N(\gamma)=1$ for $\gamma=1$.
  
There are three ranges of the distribution for $\bar{\gamma}=\min(\gamma, \xi \gamma_{\rm bo,0})$, with two breaks present at $\bar{\gamma}=\gamma_{\rm bo,0}$ and $\bar{\gamma}=\gamma_{\rm br} =\gamma_{\rm bo,0}^{1/(1-\alpha)}$. For $\alpha \ge 1$, the second break does not exist, so most of our following discussion assumes that $0<\alpha<1$.

For $\bar{\gamma}<\gamma_{\rm bo,0}$,  $\gamma<\xi \gamma_{\rm bo,0}$ in all cases, so $N(\bar{\gamma})=N(\gamma)$. For  $\bar{\gamma}>\gamma_{\rm bo,0}^{1/(1-\alpha)}$, $\xi<\xi_{\rm max}<\gamma/\gamma_{\rm bo}$, so 
$N(\bar{\gamma})=N(\xi\gamma_{\rm bo,0})=N(\xi)/\gamma_{\rm bo,0}$.  
$N(\xi)$ is given by 

\begin{align}
N(\xi)=&\gamma^{1-p} \left. \setlength\arraycolsep{1pt}
{}_2 F_1 \left(\begin{matrix}1&\frac{1-p}{\alpha} &\\
&\frac{1-p+\alpha}{\alpha}\end{matrix};\gamma^{\alpha} \right) \right| ^{\gamma=\infty}_{\gamma=\xi^{1/\alpha}} \\ \approx& \frac{p-1}{p+\alpha-1} \xi^{-(\alpha+p-1)/\alpha}\nonumber
\end{align}
where $_{2}F_1$ is the hypergeometric function. The last expression comes from an expansion of that function that applies for $\xi \gg 1$, which is the relevant limit for  $\bar{\gamma}\gg\gamma_{\rm bo,0}$.
Finally, for $\gamma_{\rm bo,0}<\bar{\gamma}<\gamma_{\rm bo,0}^{1/(1-\alpha)}$, 

\begin{align}
N(\bar{\gamma})=&N(\gamma)|_{\bar{\gamma}}+\frac{1}{\gamma_{\rm bo,0}}N(\xi )|_{\bar{\gamma}/\gamma_{\rm bo,0}}\\ &-\frac{d}{d\bar{\gamma}}\int^{\bar{\gamma}/\gamma_{\rm bo,0}}_1 \int^{\bar{\gamma}}_{\xi^{1/\alpha}} N(\gamma,\xi)  d\gamma d\xi,\nonumber
\end{align}
where the bars indicate substitution and we have made use of an identity for the distribution of $C=\min(A,B)$. Note that $N(\xi )=\int N(\gamma, \xi )d \gamma$ is the \textit {marginal} distribution of $\xi$, and is not the same as the uniform conditional distribution for $\xi$ mentioned earlier. 

Using the same expansion for the hypergeometric function, we can calculate the full distribution for $0<\alpha<1$:
{\small \begin{equation}
  N(\bar{\gamma})=N_0\begin{cases}
                               \bar{\gamma}^{-p}, &\bar{\gamma}\le \gamma_{\rm bo,0}\\
              \bar{\gamma}^{-p}(1 +\kappa \bar{\gamma}^{-\alpha}),&\gamma_{\rm bo,0}<\bar{\gamma}\le \gamma_{\rm br} \\
              \gamma_{\rm br}^{-p}(1+\kappa{\gamma_{\rm br}}^{-\alpha} )(\frac{\bar{\gamma}}{\gamma_{\rm br} }) ^{-(p+\alpha-1)/\alpha },    &   \bar{\gamma}>\gamma_{\rm br}\end{cases}\label{eq:fulldist}\end{equation}}where $\kappa=(p-1+\alpha^2-\alpha p+\alpha)/(\alpha+p-1)$ is a parameter that always falls in the range $0\le\kappa\le 1$. 

One important thing to note in the analytical distribution is the slight discontinuity at $\bar{\gamma}=\gamma_{\rm bo,0}$, where $N(\gamma_{\rm bo,0})^+ =(1+\kappa) N(\gamma_{\rm bo,0})^->N(\gamma_{\rm bo,0})^-$. This discontinuity results from the presence of particles with $\xi=1$ at all values of $\gamma$; such particles all cool down to $\bar{\gamma}=\gamma_{\rm bo,0}$, but no further. For $\gamma_{\rm bo,0}\ll \bar{\gamma}<\gamma_{\rm br}$, $N(\bar\gamma)\sim N_0 \bar{\gamma}^{-p}$. Therefore the discontinuity at $\bar{\gamma}=\gamma_{\rm bo,0}$ is only important locally. Because the electric field varies slightly with location in a realistic X-point region, there will be variation in $\gamma_{\rm bo,0}$ that smears out the distribution and eliminates this discontinuity.  As a result, we do not expect cooling to have strong effects on the particle distribution for $\bar{\gamma}<\gamma_{\rm br}$.

The typical power law index $p$ of particles in the uncooled simulation is approximately $1.75$ (see below), and our previous analysis shows that $\alpha \approx 0.4$. Therefore, we expect that $\gamma_{\rm br} \sim \gamma_{\rm bo,0}^{5/3} \gg \gamma_{\rm bo,0}$, and that the power law index above this break is about $(p+\alpha-1)/\alpha \approx 2.9$.  Therefore, cooling effects on particle acceleration in magnetic reconnection occur only very far above the burnoff limit, and they produce only a moderate break in the particle energy spectrum. 

We now discuss the validity of our analytical spectrum for real physical systems with $\gamma_{\rm bo,0}\ggg 1$. The analytical particle spectrum we have derived in this section is based on our test particle simulation results using Equation (7) for the radiation reaction force. Therefore, its results will only be valid for $B_p \ll B_c$. In the physical case for which our extrapolation is relevant, quantum effects may become important before the classical perturbation result breaks down. The more restrictive Schwinger limit  for electrons corresponds to $B_q\equiv (2/3)\alpha_{\rm FS} B_c$ and $\gamma_q\equiv(2/3)\alpha_{\rm FS} \gamma_c$.   

Because we do not know the value of $B_p$ for particles in a real physical system, we conservatively estimate that the accelerating particles experience a field $B_0$ in the plasma rest frame. Because $E<B_0$ even in the diffusion region where $E>B$, the Lorentz transformation into the frame of a particle with cooled Lorentz factor $\bar{\gamma}$ yields $B_p \lesssim B_0 \gamma$. Combining this result with Equation (\ref{eq:bpcrit}) and the definition of $\gamma_q$ then yields the condition for the validity of our classical analytical spectrum 

 \begin{equation} 
\bar{\gamma} \ll  \gamma_{q}\equiv \alpha_{FS} \gamma_{\rm bo, 0}^2\frac{2B_0}{ 3E_0}.
\end{equation}

Our main conclusion that cooling has no effect below $\gamma_{\rm br}$ is valid if $\gamma_q>\gamma_{\rm br}$.   This condition may be expressed as
\begin{equation}
\gamma_{\rm bo,0}>\left(\frac {3E_0}{2 B_0\alpha_{\rm FS}}\right)^{(1-\alpha)/(1-2\alpha)}.
\end{equation}

Taking $E_0/B_0=0.2$ and $\alpha=0.4$, this equation gives $\gamma_{\rm bo,0}>6.9 \times 10^4$, which corresponds via Equation (10) to $B_0<3.8\times 10^5$ G. Reconnection models that produce observed emission through the synchrotron mechanism in GRB prompt emission \citep{beniamini_13}, AGN flares \citep{giannios_13} and the Crab flares \citep{uzdensky_11} invoke magnetic field strengths far below this value (1 G-$10^4 $ G, $\sim$1 G, and $\sim$1 mG, respectively).  {We also note that the presence of quantum effects for magnetic fields approaching $B_q$ is likely to lessen the strength of the radiation reaction force rather than increasing it, because the classical force includes photons with energy larger than the electron rest energy whose production will be suppressed by quantum effects.} Therefore, our derived spectrum below $\gamma_{\rm br}$ should accurately estimate the effects of cooling on particles accelerated in magnetic reconnection in these systems.

\begin{figure}
\begin{center}
\includegraphics[width = 0.45\textwidth]{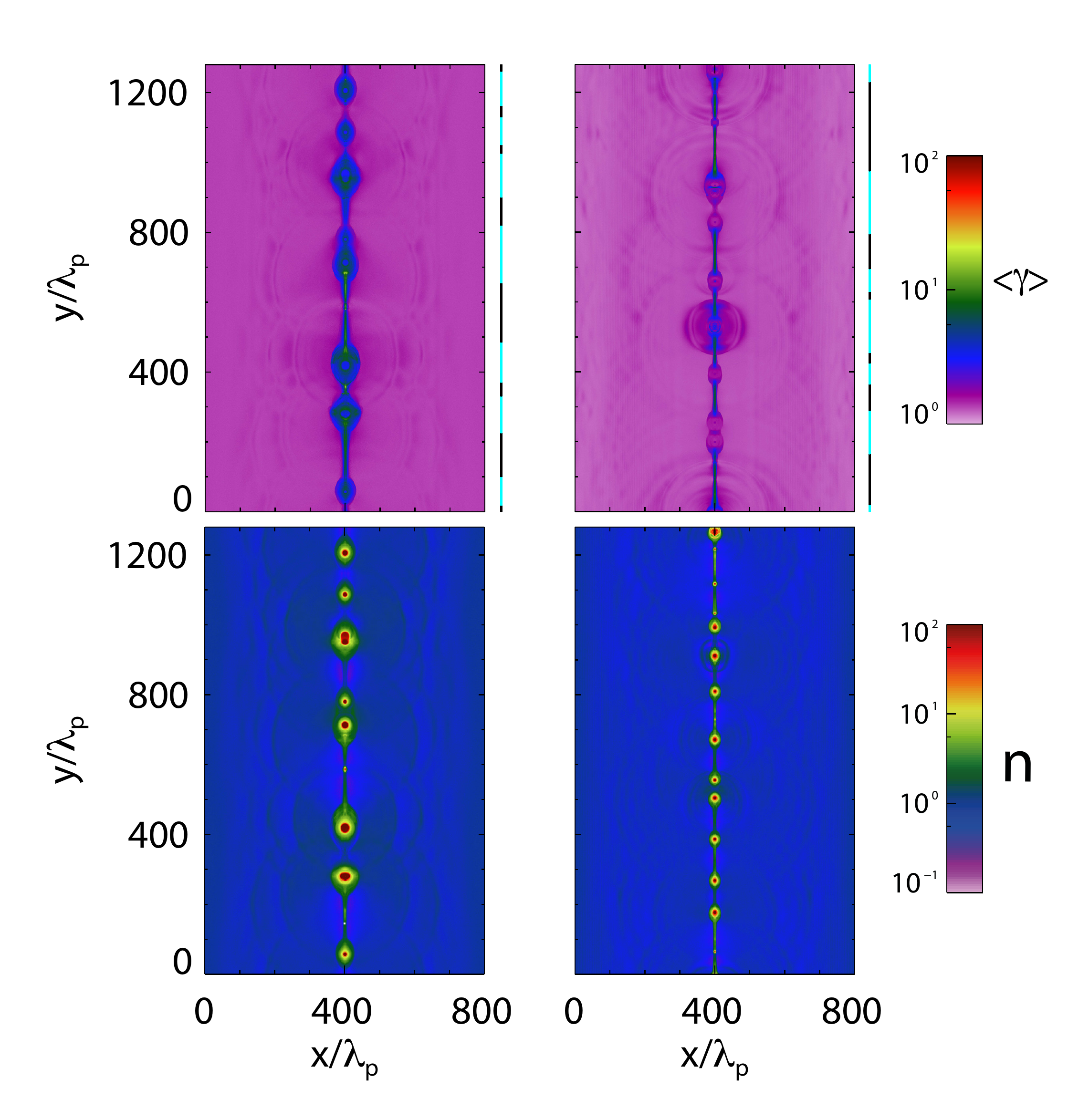}
\end{center}
\caption{ The normalized density $n$ (top) and average kinetic energy $\langle \gamma \rangle$ (bottom)  throughout the (half) simulation box at $\omegap t=562.5$ in Simulation {\tt TU} without cooling (left)  and at $\omegap t=703$ in Simulation {\tt AC} with cooling (right).  The density structures are similar in the two simulations, while bulk kinetic energies in islands and at the edges of the current sheets are significantly reduced by cooling. The black lines to the right of each box in the top row indicate the location of X-point regions within the current sheet. Fainter cyan lines indicate the other locations, which correspond to magnetic islands.  \label{fig:overallcomp}}
\end{figure}

\subsection{Effects of overall cooling on magnetic reconnection} \label{sec:comparison}
We now compare the overall structure of Simulation {\tt AC} with cooling  and Simulation {\tt TU} without cooling. We find that in Simulation {\tt AC} the evolution of the current sheet is somewhat slower, but the energy conversion rate is somewhat faster than in Simulation {\tt TU}. The global reconnection rate is approximately $0.17$,  which is $54\%$ higher than in the uncooled simulation. The typical reconnection electric field is $E_0/B_0\approx0.22$, which is $30\%$ higher than in the uncooled simulation. 

Figure \ref{fig:overallcomp} compares the densities and the average kinetic energies in the uncooled simulation {\tt TU} and the cooled simulation {\tt AC} at time $\omegap t=562.5$. The density structures of the current sheets in the two runs are similar, except for the slower evolution, as are most other properties of the reconnection.   The only significant difference is that the particles in the islands and outside of the center of the current sheet have been cooled to low energies $\langle\gamma\rangle<2$; as a result, the particle density at the very center of magnetic islands in Simulation {\tt AC} is extremely high, up to $10^4$ times the background density at the center point of islands. This cooling is in line with what we expect given that $\gamma_{\rm bo,0}=5.45$ and particles are able to cool for many plasma times if they are not being accelerated in reconnection regions.  We conclude that the structure of reconnection regions will not vary greatly due to cooling, and our test particle results may be approximately applied to uncooled simulations to estimate the effects of cooling. 
\begin{figure}
\begin{center}
\includegraphics[width = 0.48\textwidth]{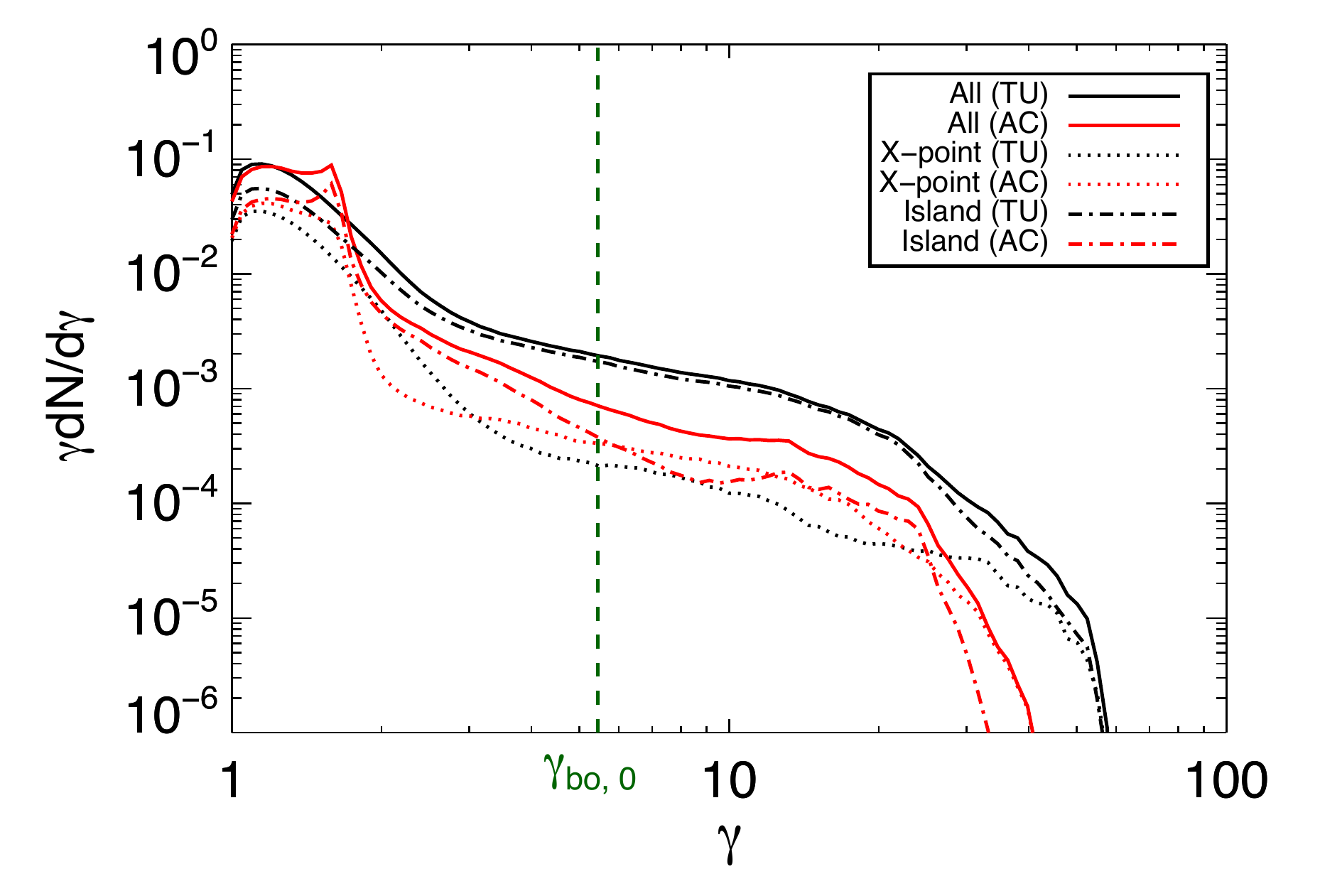}
\end{center}
\caption{ The particle energy spectrum $ dN/d \ln \gamma$ in the (half) box, within X-points, and within islands at time $\omegap t=562.5$ in Simulation {\tt TU} without cooling and in Simulation {\tt AC} with cooling. The $y$ locations of X-points and islands are shown in Figure \ref{fig:overallcomp}, while the $x$ locations span the (half) box. Spectra are normalized so that the integral of $ dN/d \ln \gamma$ over all energies is equal to 1 for the whole box in each simulation.  The sharp peak present in Simulation {\tt AC} results from the limitations on background cooling discussed in Section \ref{sec:radfeedback}. The green dashed line shows clearly that there is no cutoff at the fiducial burnoff limit $\gamma_{\rm bo,0}=5.45$ in any of the regions in Simulation {\tt AC}. The actual cutoff is located at $\gamma\sim22$ in both simulations, and is caused by their limited dynamic range.\label{fig:energyspectra}}
\end{figure}

 Figure \ref{fig:energyspectra} compares the particle energy spectra in the two simulations in the (half) box, in the X-point regions, and in the islands. The figure clearly shows that there is no cutoff at $\gamma_{\rm bo,0}=5.45$ in any of these locations.  The cutoff of the power laws is instead located at $\gamma \approx 22$ in both simulations and is a result of the saturation of acceleration due to low $\sigma$ \citep{werner_16}. For a magnetic field equal to $B_0$ this corresponds to a cutoff synchrotron energy of over 16 times the burnoff limit in the cooled simulation. The cutoff synchrotron energy may be even higher than this because the edges of islands often have a magnetic field significantly larger than $B_0$.   Thus, acceleration well beyond the burnoff limit is present even in these limited simulations. 
 
 The power law index in the uncooled simulation is $p\approx1.75$, and that in the cooled simulation is $p\approx2.3$, with no obvious break in either spectrum. In contrast, our analytical power law model with $\alpha=0.4$ predicts that because the beginning of the power law in Simulation {\tt TU} is located at $\gamma_{\rm min}\sim 3$, there will be a break in the cooled spectrum at $\gamma_{\rm br}=3(5.45/3)^{1/(1-0.4)}\sim 8.1$ with a power law index above the break of $\approx 2.9$. The disagreement between the analytical model and the actual results is not surprising given the limited dynamical range of the simulations, but it is encouraging that the slope of the power law is indeed moderately steeper in the cooled simulation.    
 
The population of particles in both the X-point regions and the islands is larger in the cooled Simulation {\tt AC} than in the uncooled simulation {\tt TU}, reflecting the difference in reconnection rate. In the uncooled Simulation {\tt TU}, the islands have far more particles than the X-points at almost all energies, with the X-points becoming important only at the highest energies of $\gamma>45$. This is because most particles eventually escape from the X-points after accelerating and enter islands without losing significant energy in the uncooled simulation, and can even undergo significant additional acceleration there \citep{guo_15}. Only at the highest energies are the X-points important, because particles that remain there are capable of accelerating for a longer time period than other particles. 

In the cooled Simulation {\tt AC}, in contrast, the X-point population makes a significant contribution to the overall energy spectrum for $\gamma>8$ and becomes dominant for $\gamma>30$. This is because the strong cooling experienced by particles in the islands greatly reduces their contribution to the high-energy particle spectrum. Nevertheless, the island particle spectrum still does not have a cutoff at $\gamma_{\rm bo,0}=5.45$, because particles are always flowing into the islands from the X-point regions as long as reconnection is ongoing, so that high-energy particles are continuously replenished there.  Overall, the burnoff limit does not limit the particle energy anywhere in the system in our fully cooled Simulation {\tt AC}, indicating that it is not an important limit at any stage of particle acceleration in magnetic reconnection. We do note that this conclusion is tentative in magnetic islands because particles there may have been affected by the presence of fields with $B_p>B_c$, but the islands do not strongly affect the overall high-energy spectrum because the island and X-point populations are comparable there. This confirms that our analytical results obtained using the TPC method are correct, and indicates that particles can be accelerated above the burnoff limit by orders of magnitude.

\section{Conclusions}
\label{sec:conclusions}
In this paper, we investigated the effect of cooling on particle acceleration in relativistic magnetic reconnection by using the TPC (test particle comparison) method, in which we compare test particle orbits in the same reconnection region with (Simulation {\tt TU}) and without (Simulation {\tt TC}) cooling. We also carried out an additional simulation (Simulation {\tt AC}) to find the effect of cooling on reconnection overall and confirm the utility of the TPC method and the parameterization of particle trajectories using the burrowing parameter $\xi$.  Our conclusions are as follows:

\begin{itemize}
\item The trajectories of the vast majority of accelerating particles are weakly affected by cooling. The most prominent effect of cooling is that the phase of oscillation across the current sheet is different for cooled particles, and cooled particles leave the X-point regions slightly earlier due to reduced inertia as a result of cooling. In rare cases, cooling causes a particle to exit the X-point region in the opposite direction from its uncooled path.

\item The acceleration of a particle in a cooled simulation can be derived from its acceleration in the uncooled simulation and its burrowing parameter $\xi$ using the prescription  $\bar{\gamma}_{\scriptscriptstyle \mathrm{f}} ={\rm min}(\gamma_{\scriptscriptstyle \mathrm{f}}, \xi\gamma_{\rm bo,0})$. Here, the burrowing parameter $\xi=\sqrt{\langle E B_0^2/(B^2 E_0)\rangle}$ indicates the extent to which the particle spends time in the current sheet and experiences a lower magnetic field (and therefore a reduced amount of cooling).

\item The burrowing parameter $\xi$ of particles varies between approximately unity  and $\xi_{\rm max}\sim \gamma_{\scriptscriptstyle \mathrm{f}}^{\alpha}$, with $\alpha=0.4$. This is roughly consistent with analytical Speiser orbit calculations, which predict that $\alpha=0.25$.

\item If the uncooled particle energy spectrum is a power law of the form $N(\gamma)\propto \gamma^{-p}$, we predict analytically that when cooling is present the distribution $N(\bar{\gamma})$ will be the same up to  $\bar{\gamma}\sim \gamma_{\rm bo,0} ^{1/(1-\alpha)}$. This is far above the burnoff limit of $\gamma_{\rm bo,0}$.  Above this break, the power law index becomes $(p+\alpha-1)/\alpha$, which is only a moderate break in the spectrum for $p <2$. The unimportance of $\gamma_{\rm bo,0}$ for the particle energy spectrum is confirmed in fully cooled Simulation {\tt AC}. 

\item In the cooled Simulation {\tt AC}, the effect of overall cooling is to speed up the rate of energy transfer relative to the uncooled simulation {\tt TU} while slightly slowing the current sheet evolution. However, there are no significant qualitative differences in the reconnection structure.

\item The particle energy spectrum in the fully cooled Simulation {\tt AC} is similar to that found in the uncooled Simulation {\tt TU}, and does not show any cutoff at $\gamma_{\rm bo,0}=5.45$. Only in the islands is the form of the spectrum significantly different in the two simulations, because that is where accelerated particles experience significant cooling. 

\item The analytical results from the TPC method are confirmed by the fully cooled Simulation {\tt AC}. We conclude that particles can be accelerated above the burnoff limit by many orders of magnitude.

\end{itemize}

Our analytical calculations of the effect of cooling on power law energy spectra, together with the measured particle energy spectra in the fully cooled PIC simulation, confirm the conclusion of \citet{cerutti_12a} that particle acceleration is not limited by cooling. Therefore, emission beyond the synchrotron burnoff limit is indeed well explained by a magnetic reconnection model.  

We now turn to the applications of our results in astrophysical contexts. We show here that reconnection is that the particle energy spectrum and the resulting radiation spectrum have no direct dependence on the burnoff limit. The TeV emission in AGN is far beyond the synchrotron burnoff limit (even accounting for relativistic boosting of the bulk of the jet). This is consistent with being produced by reconnection. On the other hand, if the Crab flares are produced by reconnection it is necessary to explain why their peak energies are only a factor of $\sim2$ above the maximum synchrotron burnoff limit of 160 MeV.  This difficulty can be resolved if the typical sizes of the flaring regions are small enough that electrons cannot reach extreme energies except in the largest flares.  Alternatively, if typical flaring regions have a strong guide field perpendicular to the plane of reconnection, most regions will not produce significant acceleration beyond 160 MeV \citep{cerutti_12a, cerutti_13a}. If future studies of the morphology of the Crab Nebula during flaring activity are consistent one of these scenarios, they will provide further evidence that high-energy flares above 160 MeV in the Crab Nebula are produced by particles accelerated in relativistic magnetic reconnection.

  \acknowledgements
 This research was  supported by the I-CORE Center for Excellence in Research in Astrophysics and by an ISA grant. TP was partially supported by a CNSF-ISF grant. DK and EN were partially supported by an ERC starting grant (GRB/SN) and an ISF grant (1277/13). 
\\ 
\\ \\

\end{document}